# Decoding Digital Influence: The Role of Social Media Behavior in Scientific Stratification Through Logistic Attribution Method

Yang Yue*

Institute of Science Studies and S&T Management, Dalian University of Technology, ysynemo@mail.edu.dlut.ed.cn

**Abstract**: [Research Background] Scientific social stratification is a classic theme in the sociology of science. The deep integration of social media has bridged the gap between scientometrics and sociology of science. This study comprehensively analyzes the impact of social media on scientific stratification and mobility, delving into the complex interplay between academic status and social media activity in the digital age. [Research Method] Innovatively, this paper employs An Explainable Logistic Attribution Analysis from a meso-level perspective to explore the correlation between social media behaviors and scientific social stratification. It examines the impact of scientists' use of social media in the digital age on scientific stratification and mobility, uniquely combining statistical methods with machine learning. This fusion effectively integrates hypothesis testing with a substantive interpretation of the contribution of independent variables to the model. [Research Conclusion] Empirical evidence demonstrates that social media promotes stratification and mobility within the scientific community, revealing a nuanced and non-linear facilitation mechanism. Social media activities positively impact scientists' status within the scientific social hierarchy to a certain extent, but beyond a specific threshold, this impact turns negative. It shows that the advent of social media has opened new channels for academic influence, transcending the limitations of traditional academic publishing, and prompting changes in scientific stratification. Additionally, the study acknowledges the limitations of its experimental design and suggests future research directions. It proposes considering a broader range of scientific fields and a more comprehensive set of indicators, building an indicator system from the perspective of scientific activity theory, while accounting for the heterogeneity of academia and inter-disciplinary differences. This research provides significant insights into understanding the distribution of power and resources in academia, as well as the diversification of scientific knowledge and innovative thinking, contributing to the healthy development of the entire scientific community system.

## 1 INTRODUCTION

The origins of social media trace back to the late 1990s, a concept initially introduced by American scholar Anthony Mayfield. However, the real surge in social media began with the establishment of MySpace in 2003. The advent of Web 2.0, emphasizing user participation and content openness, marked a shift towards a user-centered, relationship-based approach. This era empowered individuals to create and disseminate content, collectively constructing a user-centric network based on individual activities. Hence, studying the stratification and mobility within the scientific community in this social media era is of paramount importance [1]. Modern science has evolved into a vast, organized, socialized, and complex system of activities, fundamentally based on scientific endeavors [2]. Broadly speaking, scientific activity encompasses various actions involved in the scientific research process, serving as the bedrock for the sustenance and development of the scientific community [3]. Regarded as an advanced social activity for knowledge production, the complexity of scientific activities has intensified in the era of social media [4]. Starting from scientific endeavors, exploring the status and influence of scientists in the scientific community during the social media era presents significant challenges. The stratification and mobility within the scientific community are classic issues in the sociology of science [5]. Before the rise of social media, the influence and reputation of scientists were primarily based on academic achievements and peer evaluations [6]. However, the multifaceted impact of social media marks a shift from closed to open systems [7]. Its widespread adoption has made scientific activities more interwoven with social, cultural, and economic spheres, revealing limitations in traditional scientometrics indicators to fully capture the cross-dimensional impacts of scientific activities [8]. Social media provides a broader and more immediate platform for information dissemination, potentially affecting scientists' reputations and influence based on social media engagement [9]. Despite the growing visibility of social media's impact in the scientific world, research on how social media engagement and the stratification structure of the scientific community influence each other remains scarce.

The influence of scientists on social media is determined not only by their academic achievements but also by their activity and content appeal on these platforms [10]. Additionally, real-world status, prestige, reputation, and resources may be mirrored to some extent on social media, further affecting scientists' social media engagement [11]. Social media is reshaping the hierarchical structure and power distribution within the scientific community, necessitating adjustments to traditional academic evaluation systems [12]. Altimetric, as a measure of academic social media attention, has been formally integrated into technology assessment metrics. Alongside traditional citation indicators, Altmetrics are now applied in fund applications, academic promotions, result evaluations, and research management, scrutinizing their reliability and utility by researchers, fund managers, higher education institutions, and even government decision-making departments [13]. This has led to the acceptance of Altmetrics as an indicator of influence, supplementing traditional metrics which have quantitative blind spots. Traditional quantitative indicators, like citation counts, impact factors, and h-indices, show clear inadequacies in the context of social media. Especially, they fail to fully reflect the multidimensional cross-impacts of scientific activities on society, culture, and economics. This unidimensional evaluation perspective greatly limits the understanding of the complete landscape of scientific activities [14].

From a meso-theoretical standpoint, this paper explores under what conditions social media engagement can serve as an indicator of scientific recognition. We offer a new perspective to understand and contrast the extent of scientific recognition in the digital and real worlds, attributed to the varying influence of scientific work [15]. Commonly, it is believed that influence on social media relies more on personal activity and the appeal of scientific communication, whereas real-world influence hinges more on academic achievements and status. However, this is not entirely the case [16]. On social media platforms, in addition to personal activity and the attractiveness of shared scientific content, a scientist's real-world reputation and scientific recognition still reflect in the digital realm. This mapping is a many-to-one relationship, making it challenging to discern the individual contributions of different factors to social media engagement. The engagement on social media represents a composite of various aspects. This paper aims to explore whether social media impacts the social stratification in the scientific community and affects the mobility within this stratification. We primarily focus on the changes in the scientific community's social hierarchy post the advent of social networks and the extent of these changes to comprehend the impact of social media on scientific social stratification. The emergence of social media has significantly altered the modes of information dissemination, potentially leading to changes in scientists' influence and prestige. By comparing the levels of scientific social stratification before and after the rise of social media, we can explore the overall impact of social media on the stratification structure of the scientific community and understand how this new mode of information dissemination and interactive communication is changing the power structure within the scientific world.

## 2 LITERATURE REVIEW: MULTIDIMENSIONAL PERSPECTIVES ON STRATIFICATION IN THE SCIENTIFIC COMMUNITY

### 2.1 Stratification in the Scientific Community and Academic Prestige

In social stratification theory, prestige often extends beyond specific groups. Broadly, it encompasses a wider spectrum [17], covering public, organizational institutions [18], and third-party evaluations [19]. From this perspective, the extent of social media engagement closely aligns with the concept of prestige, serving as an effective tool for measuring and representing it [20]. However, in the realm of the scientific community, prestige is specifically regarded as academic prestige. Due to the unique nature of the scientific community, prestige stratification is often narrowly interpreted as the hierarchical arrangement of academic prestige. The universalism and fairness inherent in the scientific community imply that the status, opportunities, and rewards scientists achieve typically correspond to their income, prestige, and influence hierarchy. In such a context, social media engagement inevitably impacts the stratification of scientific prestige, primarily manifested in the following two aspects:

- Firstly, the use of social media is deeply integrated with scientific activities. Social media breaks down barriers of time and space, expanding the range and possibility of communication among researchers, thereby enhancing their prestige. By constructing their online behavior and shaping their digital image, scientists create a unique academic influence in the virtual world. Continuously enhancing their comprehensive academic influence online can impact their real-world persona, elevating their academic prestige in the real world. Previous studies have shown that higher social media engagement can enhance a scholar's academic prestige. For instance, authors with higher social media engagement enjoy greater exposure, becoming more recognized by peers and other scholars. If the exposure is positive, it contributes to a positive accumulation of academic prestige.
- Secondly, social media engagement may have a direct or indirect impact on a scientist's academic prestige. It provides a platform for researchers to display and share their research findings, opinions, and information, thereby attracting attention and feedback. This interaction and level of engagement can intuitively reflect a researcher's prestige among peers and the public. Social media facilitates both scholar-to-scholar and scholar-public-scholar communication models, extending the influence beyond academia to a broader public. This significantly increases the reach and impact of academic work, unmatched by traditional scientific communication methods. The real-world academic prestige hierarchy is likely to be influenced by social media engagement. On social media, scientists can accumulate followers and increase their visibility, thereby enhancing their academic prestige. Therefore, scholars with higher social media engagement may positively impact their academic prestige elevation.

### 2.2 Perspectives on Influence within the Scientific Social Stratification

Influence is a complex concept, involving an individual's, group's, or institution's capacity to affect and control others' behavior [21]. Influence is understood as the exerted force by an influencer upon the influenced, existing not only between individuals but also among groups or institutions [22]. The exertion of influence requires the motivation and capability of the person or organization, along with the support for their actions [23]. The core of influence is the ability to control, manifesting as the influencer's purposeful control over the recipient's cognition, tendencies, opinions, attitudes, beliefs, and outward behavior [24]. Influence can be broadly defined as the capacity of a person or entity to affect the behavior, thoughts, or decisions of others. This influence can be direct or indirect [25], positive or negative. In a narrower sense, influence refers to the capacity to affect others within a specific field or environment, such as social media influence or a scientist's academic influence [26].

Within social network models, influence is considered the ability of an entity or group to affect the behavior, attitudes, or views of others in the network [27]. This influence can be exerted through word-of-mouth, social media sharing, or direct interpersonal interactions. Thus, the concept of social media influence emerges: it refers to the influence gained through online and social media platforms, distinct from traditional pathways of authority and status acquisition. It is measured by metrics such as follower counts, likes, shares, and comments. This form of influence can spread rapidly and have a wide reach but may also be fleeting. On social media, academic and social influence together constitute social media influence. In the scientific field, social media influence merges categories of academic and social influence, representing not only a scientist's online influence but also a reflection of their real-world influence. Social media platforms, through selective information dissemination, can influence users' opinions and behaviors, thus generating influence. This influence depends not only on the content of the information but also on the entity disseminating it, the mode of dissemination, and the environment. In the age of social media, the concept of influence is also closely linked to the network's out-degree. In social network models, influence is described as the number of outward links a node sends; the more links, the greater the influence.

In various fields, the definition of influence varies based on different scholars' perspectives, research questions, contexts, and objectives. For instance, influence can be categorized differently based on the subject and audience, such as personal influence [28] and social influence. If we differentiate influence based on the online and offline realms, we can divide it into digital influence [29] and real-world influence. In the digital society, personal influence is often reflected through metrics like follower count, likes, and comments. However, in the real world, personal and social influence are often manifested through organizational mediums, reflecting the capacity to impact others and society as a whole [30]. In practice, the definition of influence is quite vague across different applications. Regardless of the field, influence is commonly viewed as a valuable resource that helps individuals gain more reputation, opportunities, and resources within a specific domain [31]. Sociologists often define individual influence as a person's fame, reputation, and ability to influence, considering it a form of power capable of shaping opinions, decisions, and actions [32]. In the scientific field, the concept of influence has its distinct characteristics [33]. Scientific influence is usually measured by the quality and quantity of academic achievements, as well as the frequency with which these works are cited by other scholars [34]. The more citations a paper receives, the greater its influence in the relevant field. This measurement of influence is often closely related to a scientist's academic prestige and status [35].

The perspective that scientific social stratification is manifested through academic influence suggests that academic influence is typically measured by factors such as research quality and quantity, citation count, and academic reputation [36]. Academic influence can include the impact of scholarly papers and the influence of scholars themselves, as these forms of influence are primarily disseminated through academic publications and recognition within the academic community [37]. Referring to Merton's definition of the scientific community, a scientist's position within this community is determined by their actual contributions to science [38], not just quantifiable contributions. It's crucial to understand the distinction between metrics and peer review. The current prevalent approach conflates academic influence with social influence, despite their clear differences.

Traditionally, in academic settings, the Impact Factor (IF) has been commonly used to represent academic influence [39]. However, given the numerous issues with using the Impact Factor to characterize academic influence, there is now a growing preference for employing the H-index as a measure of academic impact [40]. This preference arises because the H-index effectively addresses issues while aligning with Merton's definition of scientific contributions. The appeal of the H-index lies in its ability to highlight researchers who have made lasting and significant contributions but have not received rewards commensurate with their reputation. This phenomenon is widespread and illustrates the mismatch between influence and scientific recognition [41]. It's important to note that influence is dynamic. Therefore, when discussing influence, the impact of time must be explicitly considered. The reluctance to use the Impact Factor (IF) as an indicator of academic influence stems from the fact that influence changes over time; it can increase or decrease. A scholar's influence is not static; it can grow or diminish. Therefore, when using citation counts as a measure of academic influence, this variability should be taken into account [42]. Consequently, tracking the changes in a scholar's H-index throughout their academic career and dynamically assessing their influence is an important exploratory approach.

### 2.3 The Essence of Scientific Social Stratification: Unifying Academic Influence and Scientific Recognition

Scientific recognition refers to the acknowledgment and acceptance of a scientist by the scientific community, universities, and research institutions. A scientist's status within the scientific community is built upon this recognition. Merton introduced the concept of "scientific recognition," indicating the degree to which the scientific community accepts and acknowledges scientific achievements, scientists, and scientific issues. He posited that scientific recognition is a form of social stratification within the scientific world, where scientists must consider not only the truth of science itself but also the level of acceptance and acknowledgment by their peers. Merton believed that gaining and maintaining scientific recognition requires adherence to certain norms and values of the scientific community, such as fairness, objectivity, replicability in research, and high levels of expertise and skills among scientists [43]. The acquisition of scientific recognition also involves factors related to social context and the relationship with scientific institutions. Overall, Merton's proposition of "scientific recognition" emphasizes the influence of social and cultural backgrounds on scientific research [44] and is of great significance in understanding scientific development and policy-making [45].

This paper further expands and enriches the concept of scientific recognition in the era of social media. In this context, scientific recognition on social media is seen as social media influence stripped of societal influence (i.e., social media scientific recognition = social media influence - societal influence). The widespread use of social media has changed the traditional mechanisms of scientific recognition. In this new era, scientific recognition is no longer solely based on the quality and innovation of academic research but also integrates activity and public impact on social media [46]. The intertwining of academic and social influence of scientists forms the new standards affecting the stratification of the scientific community. Thus, scientific recognition has evolved from a singular academic evaluation criterion to a complex system incorporating multiple dimensions [47]. In the era deeply shaped by social media, the standards and processes of scientific social stratification are undergoing significant transformations. The traditional stratification mechanism, primarily driven by academic achievements and research contributions, is evolving into a more comprehensive evaluation system, where a scientist's activity and public influence on social media play a crucial role. This new stratification standard not only emphasizes the importance of academic achievements but also focuses on the frequency and quality of social media interactions and the impact on a broader audience. Such a multidimensional evaluation mechanism makes scientific social stratification more aligned with the realities of the modern scientific community, adding dynamism and diversity.

Due to the complex nature of influence, defining it and selecting appropriate indicators can be challenging. This paper introduces the concept of scientific recognition, explicitly defining the impactful components in the process of influence dissemination. Scientific recognition affirms the influence of scientific work, representing acknowledgment and reward for a scientist's achievements. It is a broad term encompassing various forms, including honors, prestige, follower counts, awards, and professional titles. The scientific system has developed a reward system to confer recognition and respect to scientists who make genuinely original contributions. Forms of scientific recognition mainly include (1) honors and awards, (2) funding support, (3) professional positions, and (4) peer acknowledgment. Honors and awards are common forms, including the conferment of awards and memberships. Ideally, these honors increase a scientist's visibility, establish a positive image, praise their work, and encourage original contributions [48]. Merton observed that scientists at different levels of the scientific stratification system are variably motivated by prestigious honors and awards. For scientists, awards are not only a testament to successful work but also increase the likelihood of securing more resources [49]. Such honors often come with monetary support. However, recognition through awards is limited, as only a few scientists can achieve acknowledgment in this way [50].

Another form of recognition is funding support, where original and contributive work is more likely to secure funding. A third form involves holding prestigious positions in associations, universities, and research institutions, serving as a reward for outstanding scientists. Additionally, the fourth form of scientific recognition, as termed by Merton, is renown, measured using citation indexes but essentially representing peer acknowledgment. It reflects the type and degree of attention from the scientific community, serving as an indicator of feedback from the community. In traditional scientific activities, citation behavior is considered a manifestation of renown.

Scientific rewards are a key component of the scientific recognition system. Ideally, they should prioritize researchers who have made long-term and significant contributions to academia but have not received recognition and rewards commensurate with their achievements. However, such instances are relatively rare, and the discrepancy between academic influence and received scientific recognition is increasingly evident [51]. This misalignment between academic impact and scientific recognition in academia is the primary cause of this issue. Before the professionalization of science, the system was simpler, and scientific contributions were more readily acknowledged by peers, with less deviation between academic influence and scientific recognition [52]. During this period, scientific influence could essentially be equated with scientific recognition.

However, with the professionalization and institutional expansion of science, along with varying evaluation standards and cultural environments, a significant divergence has emerged between academic influence and scientific recognition. During this era, it has become challenging to maintain consistency between these two, leading to disparities. This divergence can be understood in terms of the in-degree and out-degree relationship in social networks. In an ideal environment, academic influence is equal to scientific recognition. There is some overlap between influence and recognition, but they are fundamentally different, as influence does not guarantee complete recognition. Analyzing the concept of influence reveals a certain relationship between influence and scientific recognition. In the age of social media, the concept of scientific recognition can be interpreted as the in-degree in social network models, where recognition is depicted by the number of links received by an entity node. The more links, the stronger the degree of scientific recognition, and these factors together constitute the complex phenomena of the scientific community. The limited nature of reward resources contributes to the observed disparities, such as researchers' excessive pursuit of short-term results and overemphasis on trending topics.

Additionally, the multidimensional perspective of scientific social stratification research is reflected in its interdisciplinary nature. Recent years have seen an emergence of numerous outstanding studies addressing various aspects related to scientific social stratification. For example, Bihagen et al. explored gender differences in occupational maturity in Sweden [53], while Light focused on the intrinsic relationship between gender stratification and publishing in American science [54]. González-Sauri et al. investigated the influence of early-career academic prestige stratification on scholars' future academic achievements [55]. Furthermore, Posselt et al. discussed the relationship between graduate education and social stratification [56], and Canche delved into the salary disparities among community college scientists [57]. In terms of gender and racial stratification in the scientific community, Fox et al. concentrated on the relationship between women, gender, and technology [58]. Ecklund et al. explored gender segregation in elite academic science [59], while Sá et al. studied the gender gap in elite scientists' research productivity and recognition [60]. Thomson et al. revealed pathways of reproducing racial-ethnic stratification in American academic science [61]. In the realm of scientific publishing, Sugimoto et al. uncovered age stratification and cohort effects in scholarly communication [62], and Siler et al. focused on author and institutional stratification issues in open-access publishing [63]. Carayol et al. analyzed novelty, impact, and journal stratification in science [64]. Higher education is a significant field of study for scientific social stratification; Bloch et al. revisited status stratification in higher education reforms [65], complemented by Collins's study offering comparative principles [66]. Yeung explored the expansion of higher education and social stratification in China [67], while Davies et al. studied structural inequalities in Canadian and American universities [68]. Marginson focused on the global stratification of higher education at a macro level [69]. Regarding the interaction of science, technology, and society, Chinn et al. found a trend of scientists being perceived as social threats [70], while Falkenberg et al. researched the cognitive consequences of technological innovation and capitalization in academia [71]. Francis et al. explored the impact of scientific stratification on the UK school curriculum [72]. In the realm of research collaboration and internationalization, Faria et al. examined cooperative dynamics in American transplant research [73], and Kristensen investigated the international relational networks of American social science researchers [74]. Hoekman analyzed proximity and stratification in European research collaboration networks from a policy perspective [75], and Shu et al. studied the institutionalized stratification of China's higher education system [76].

**2.4 Theoretical hypothesis**

Drawing on Merton's middle-range theory, this paper empirically designs and hypothesizes the relationship between users' social media usage and scientific social stratification from a meso-level perspective. Stratification and mobility are classic issues in sociological research. Studying these phenomena in the era of social media at the meso level is of significant importance. This chapter demonstrates the changes in scientists' levels within the scientific social hierarchy post the advent of social media at the meso-behavioral level and measures the degree of change in scientific social stratification. It is important to note that this stratification is not a strict categorization based on specific indicators, but rather a portrayal and description of researchers at the meso-behavioral level.

Within the framework of scientific activity theory, open social media networks are seen as a multi-layered, multi-dimensional scientific data system encompassing citation data, forwarding data, automatic data, and user-generated content, forming an open data ecosystem [77]. Utilizing the scientific activity theory to comprehensively interpret the scientific data and its multidimensional impact within social media networks offers a holistic understanding from both scientific and sociological perspectives. This paper aims to provide theoretical support for science policy formulation, the development of the scientific system, and the cultural construction of the scientific community through such research.

Focusing on the field of scientometrics, this paper compares the state and changes in scientific social stratification before and after the emergence of social media. It posits related hypotheses to explore the impact of social media on the hierarchical mobility within the scientific community and further analyzes the influencing factors of social media usage on scientific social stratification and mobility, as well as the extent of its impact on the structure of scientific social stratification. It also examines under different conditions whether such impacts stem from academic influence on social media or real-world influence.

- This paper explores the impact of social media on the stratification of the scientific community. Social media has transformed the way information is disseminated, potentially influencing scientists' impact and prestige. Given the noticeable cumulative effect of social media, a proactive approach may play a significant role. Scientists who engage more with their peers on social networks are likely to benefit more than those with passive accounts. Here, user behavior is categorized into active and passive types. Based on this, we hypothesize that an active behavioral character significantly and meaningfully influences one's position in the scientific social hierarchy. Therefore, we propose the null hypothesis。
H0: Active behavior on social media does not significantly influence a scientist's position in the scientific social hierarchy. The alternative hypothesis H1: Active behavior on social media significantly influences a scientist's position in the scientific social hierarchy.

- The widespread use of social media plays a crucial role in academia. The habitual use of social media by scholars can impact their academic status. On social media, scientists and scientific topics are more likely to garner attention and discussion from peers, allowing practitioners at different levels to exert their influence more easily. This phenomenon could further intensify the stratification of the scientific community. A scholar's social media habits could significantly influence their academic status, and effective social media engagement may bring greater benefits to scholars. This study aims to explore how social media influences the stratification structure of the scientific community and understand how this new type of social media behavior changes the power dynamics within the scientific world. A scholar's involvement on social media impacts their stratification status; the more active they are on social media, the greater the impact on their stratification status. A scientist's activities on social media platforms increase their visibility and attract more attention and recognition from their peers, thereby enhancing their influence and affecting their level of mobility and stratification. Therefore, we propose the hypothesis that a scientist's level of activity on social media platforms influences their mobility within the scientific social hierarchy. The higher the activity indicators on social media, the greater the level of scientific social stratification and mobility.Null hypothesis H2: A scientist's level of activity on social media platforms has no significant correlation with their position in the scientific social hierarchy. Alternative hypothesis H3: A scientist's level of activity on social media platforms has a significant correlation with their position in the scientific social hierarchy. Empirical research methods are designed to test these hypotheses and empirically verify the proposed assumptions.

## 3 RESEARCH METHODOLOGY AND EXPERIMENTAL DESIGN

### 3.1 Empirical Research Design

Methodology: This paper utilizes an interpretable logistic attribution algorithm. In the first half of the paper, traditional statistical methods are employed for statistical analysis: using logistic regression for hypothesis testing and estimating the impact of independent variables on the dependent variable. Hypotheses are first proposed, followed by an attribution analysis of the data from independent variables to the dependent variable. By examining the P-values, the study determines which independent variables have a statistically significant (or insignificant) impact on the dependent variable, thereby validating the proposed hypotheses. In the latter half of the paper, from a machine learning perspective, logistic regression is used as a binary classification algorithm to assess model performance and variable importance [78]. The constructed logistic regression model is evaluated and interpreted using SHAP (SHapley Additive exPlanations) values to analyze the contribution of each independent variable to the overall model. SHAP is used to assess the contribution of variables to the model and the importance of various factors affecting scientific social stratification and mobility. This reveals the phenomenon of scientific social stratification in the era of social media and the changes in the social hierarchy of the scientific community in the social media environment.

Research Design: Firstly, establish rules to separate academic influence from social media engagement. When using social media engagement, which includes both academic and social influence, for measuring academic influence, it's crucial to differentiate the two. Especially when using social media engagement as an alternative indicator for scientific social stratification, the following basic rules should be followed: under specific conditions, social media engagement should be more indicative of academic influence rather than social influence. Rules based on the characteristics of social media are formulated. Rule 1: Social media accounts must have a clear professional declaration; Rule 2: Social media accounts should be dedicated to scientific communication and dissemination, excluding influence gained from other factors. When these two conditions are met, social media engagement in the network is considered representative of the author's academic influence. Thus, under these conditions, the level of scientific social stratification on social media may reflect a scientist's influence and recognition within the social media environment. This will measure the mapping relationship between real-world and online scientific social stratification, determining if there is a change in stratification and the trend of such change.

Secondly, this paper uses 2008 as a watershed year to analyze the impact of social media on the scientific social stratification of scholars. Although social media emerged in 2006, its impact became more pronounced after 2008, as shown in Figure 1. Data analysis on X account (formerly Twitter), including posts, followers, and time distribution, supports the selection of 2008 as the year when social media was widely used in the scientific field. The data indicates that scientists became generally active on social media around 2009, with influential accounts being established between 2009-2012. The study selects indicators from two periods: the traditional H-index and the emerging Altmetrics indicator. After 2010, the incorporation of Altmetrics into scientometrics became a trend, reflecting the different standards for measuring scientific social stratification before and after the advent of social media. The H-index, as an early significant indicator, reflects a scholar's status in the scientific community, while Altmetrics, emerging after 2010, represents the new trend in scientometric indicators in the era of social media. The number of Followers of X accounts (formerly Twitter) that meet the criteria is chosen as an alternative metric for scientific social stratification in the era of social media. Considering the fundamental differences between the two indicators, this study uses the method of ranking percentage difference to quantify the extent of change in scholars' scientific recognition before and after the advent of social media, thereby reflecting the changes in scientists' stratification within the scientific community at the meso-behavioral level.

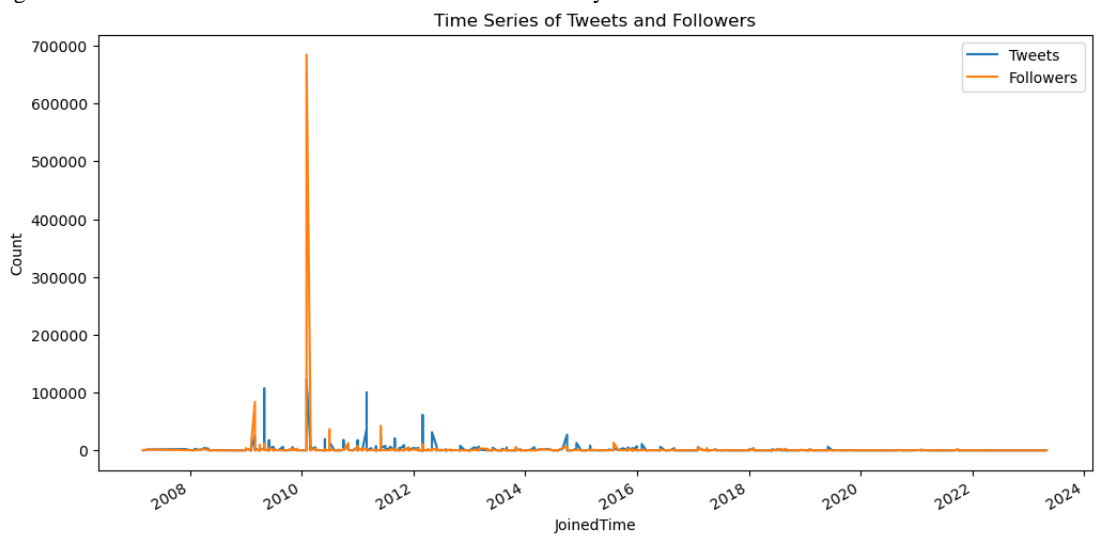

Figure 1: Time series of account creation and activity

Finally, big data technology is utilized to construct social media user profiles and analyze these data to study scholars' social media behaviors and the potential impact of social media on the mobility within the scientific social stratification. Data collection involves using social media crawling techniques to gather relevant data from the verified accounts of authors, including the number of posts, account existence days, follower counts, following counts, etc. User profile construction involves creating indicators to analyze the crawled data, depicting scholars' social media behavior characteristics. This may include active times on social media, post density, follower growth rate, following ratio, and Composite Activity. These indicators help describe scholars' activity levels, influence growth, social strategies, and overall performance on social media. Lastly, scientific social stratification mobility is defined as the dependent variable, with social media behavior indicators as independent variables. A logistic regression model is employed to verify the correlation between social media behavior indicators and scientific social stratification mobility. Analysis of the logistic regression results, including the significance of coefficients and the model's fit, is conducted to test hypotheses and explain the relationship between social media usage habits and mobility within the scientific social stratification.

### 3.2 Data Source and Acquisition

The study targets authors in the field of scientometrics, focusing on two journals: Scientometrics and Informetrics. The selection process involves filtering authors, choosing core corresponding authors from the directed author collaboration network as the study subjects. Around this group of authors, two types of data are collected: scientometric data from the Web of Science website, focusing on these two journals due to their high popularity and acceptance among scholars in scientometrics, particularly in the Altmetrics research field. The other type of data involves manually verifying the X accounts (formerly Twitter) of the core corresponding authors that meet the previously mentioned rules. The social media data of scientists on X accounts (formerly Twitter) are collected through APIs and manually verified. All data was gathered on June 1, 2021, and updated on June 1, 2023.

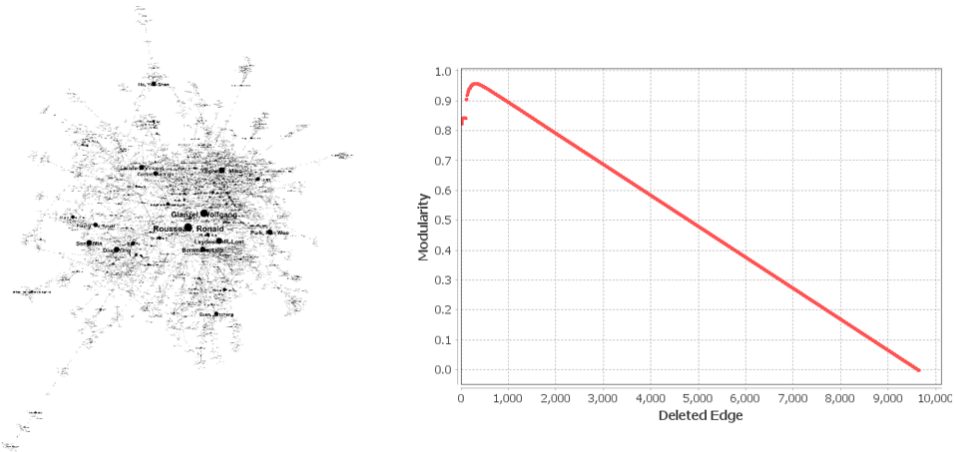

Figure 2: Core authors identified based on the Girvan-Newman community detection algorithm

The process involves filtering the sample data. The theoretical basis for selecting key core researchers: Typically, author collaboration networks are undirected graphs. However, in the highly stratified scientific community, the relationships in collaborations are unequal, often showing resource bias. Namely, the corresponding authors within a group, as significant members, bear more responsibility and provide more resources. Therefore, the author collaboration network, from the perspective of scientific social stratification, is a directed graph. In this directed graph, corresponding authors, as key nodes, have representative significance for the description of scientific social stratification. However, not all corresponding authors can provide effective resources. Significant influence within the group of corresponding authors becomes the sample target for this study, as this group is crucial for researching scientific social stratification.

The process of selecting these corresponding authors is critical. The study uses community detection algorithms for selection and proposes the following criteria: By analyzing the collaboration network of authors from the two journals and applying community structure algorithms, key community structures within the collaboration network are detected. The Girvan-Newman algorithm is used to analyze the author collaboration network of these two journals. The network consists of 1020 nodes and 9632 edges. Using the Girvan-Newman algorithm, the Maximum found modularity coefficient is calculated to be 0.96, indicating high community structure characteristics, as shown in Figure 2. The number of detected communities is 1796. These 1796 nodes are matched with corresponding authors, resulting in 8551 corresponding authors, and after deduplication, 3927 individuals are identified. Matching results in 1276 corresponding authors. Based on this, the X accounts (formerly Twitter) of these scientists are queried through the API, and the accounts meeting the experimental rules 1 and 2 and making professional declarations are manually verified. A total of 459 valid data points are obtained, and this group of authors serves as the target sample for this chapter's study.

### 3.3 Variable Definition and Measurement

(1) Dependent Variable: This study identifies the degree of change in scientific social stratification before and after the emergence of social media as the dependent variable. Social media user profile indicators and scientific activity indicators of scientists are used as explanatory variables, where the dependent variable is a binary one, indicating either an increase or decrease in status.

(2) Independent Variables: The study is positioned in the field of scientometrics, focusing on core corresponding authors. Based on this, social media user profiles and influence indicators are constructed, including (1) account joining time, (2) post density, (3) follower growth rate, (4) following ratio, and (5) Composite Activity. These indicators are critical in the industrial operation of social media. Post density measures the frequency of posts over a specific time, reflecting the user's activity level on social media. This indicator is essential for depicting the user's state on social media.

$$\text{Post Density} = \frac{\text{Number of Posts}}{\text{Account Existence Days}}$$

The follower growth rate measures the user's ability to attract new followers, describing the user's growth over a period. Due to data acquisition challenges, this study uses the overall growth rate to assess the user's overall level of influence.

$$\text{Follower Growth Rate} = \frac{\text{Current Followers} - \text{Historical Followers}}{\text{Time Interval}}$$

The following ratio reflects the user's position in the social network. A high ratio indicates an active strategy in social media usage, differentiating active and passive behavioral types.

$$\text{Following Ratio} = \frac{\text{Number of Followed Accounts}}{\text{Number of Followers}}$$

The Composite Activity combines post frequency and social network participation, reflecting the user's comprehensive activity level on social media.

$$\text{Composite Activity} = \alpha * \text{Post Density} + \beta * \left(\frac{\text{Number of Followers}}{\text{Number of Followed Accounts}}\right)$$

Additionally, scientific activity indicators include the number of publications, citation count, average citation per publication, and weight of authorship in the journals. These traditional scientometric indicators are used as independent variables. Since a scholar's influence originates from their scientific contributions in reality, social media and scientific activity indicators are used jointly as independent variables, not as control variables. Notably, considering stratification mobility indicators originate from the H-index, the H-index is not used as an independent variable to avoid multicollinearity.

The logistic regression model will assess the relationship between social media usage habits and scientific social stratification mobility. The model will examine the significance of coefficients, model fit, and other relevant statistics to test hypotheses and interpret the relationship between social media usage and scientific social stratification mobility.

## 4 RESULTS

### 4.1 Descriptive Statistics

First, a descriptive analysis of the journal authors is conducted. Using Google Maps API, the geographic coordinates of the corresponding authors' addresses are encoded and visualized, successfully identifying the leading countries in scientific research and the global distribution density of scientific studies, as shown in Figure 3. The results reveal that their influence has radiated to every corner of the world. The top five countries in scientometrics are China, the United States, Spain, Germany, and Brazil, covering several major language systems. This reflects not only the global prevalence of scientometrics research but also the broad representativeness of choosing this field for studying scientific social stratification. From a geographic distribution perspective, Europe, as the birthplace of scientometrics, has the highest research density, followed by North America and Asia.

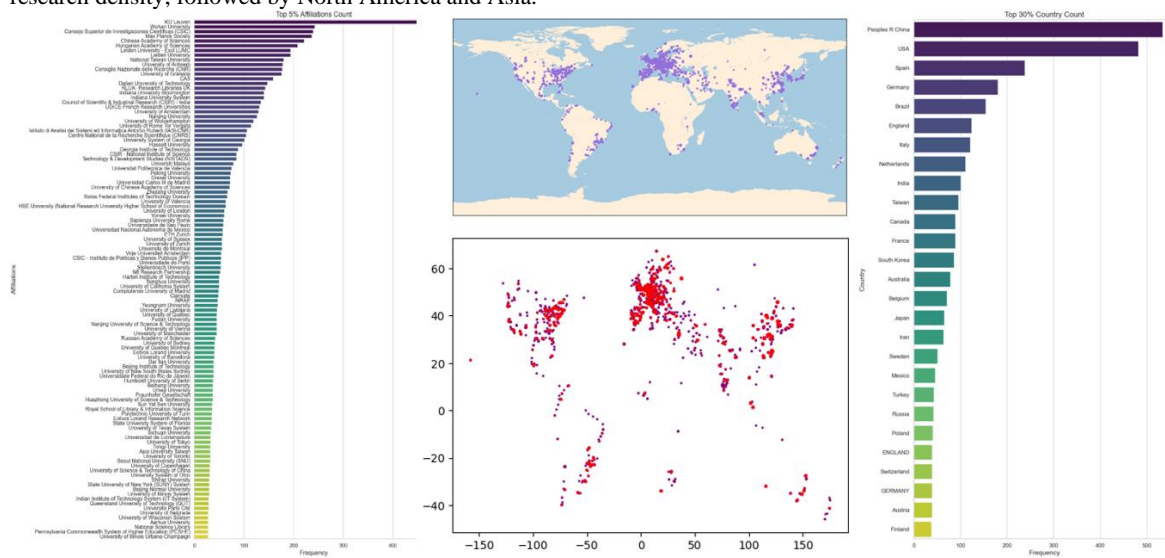

Figure 3: Distribution of scientometrics fields

A descriptive statistical analysis of the overall data is performed, as shown in Table 1, presenting the descriptive statistics of the main variables. The results indicate significant individual differences among the scientific community in terms of social media and scientific activity indicators. These differences may be related to individual social media strategies and are also influenced by their scientific research behavior and their status and influence in the scientific community. A deeper analysis of these data shows that most variables exhibit a significant skewness in distribution, indicating extremes in many variables. The skewed distribution of characteristics also suggests the presence of some very active or important users. Regarding the account registration time, the average is 3476.9 days with a standard deviation of 1511.2 days, indicating significant differences in the authors' recognition and sensitivity to social media. A longer account registration time might imply early active attention and use of social media. The following ratio (FR) has an average of 2.1 and a standard deviation of 4.18, showing significant differences in users' strategies and behavioral characteristics in social networks. Some users tend to establish more connections, while others may act more as information receivers. Other indicators like post density (TD), follower growth rate (FGR), number of publications (P), number of citations (C), average citation per publication (PC), and number of appearances in journals (AW) also reveal considerable individual differences among authors. The Composite Activity (CA) has an average of 5.68 and a standard deviation of 53.92, indicating vast differences in users' activity levels on social media. Some users may be very active, while others are relatively passive.

Table 1: Descriptive Statistical Analysis of Variables

| Variables | Mean | Standard Deviation | Minimum | Median | Maximum |
|---|---|---|---|---|---|
| AD | 3476.904 | 3476.904 | 222 | 3782 | 6127 |
| TD | 1.207063 | 1.207063 | 0 | 0.065186 | 125.1564 |
| FGR | 0.650288 | 0.650288 | 0 | 0.057051 | 112.336 |
| FR | 2.100178 | 2.100178 | 0 | 0.974026 | 59 |
| CA | 5.683231 | 5.683231 | 0 | 0.578615 | 1060.379 |
| P | 21.34205 | 21.34205 | 0 | 4 | 669 |
| C | 626.0697 | 626.0697 | 0 | 42 | 30118 |
| PC | 16.64327 | 16.64327 | 0 | 1.25 | 392.29 |
| AW | 5.62963 | 5.62963 | 2 | 3 | 120 |

Overall, while certain variables exhibit linear relationships, most show weak correlations. A correlation analysis and Variance Inflation Factor (VIF) analysis reveal that account registration time (AD) generally has a low correlation with other variables, indicating a non-significant link between registration time and other social media activity or scientific activity indicators. This may imply that authors' recognition of social media does not directly relate to their activity level or scientific achievements on these platforms. Post density (TD) has a moderate positive correlation with follower growth rate (FGR) and Composite Activity (CA), suggesting that accounts with higher post density tend to attract new followers more effectively and are more active on social media. Follower growth rate (FGR) is highly correlated with Composite Activity (CA), indicating that users with high influence and content appeal are typically more active on social media. The following ratio (FR) has low correlations with other indicators, suggesting that users' social strategies (active or passive) do not significantly link to their activity level or scientific indicators on social media. Composite Activity (CA) is highly correlated with follower growth rate (FGR), reflecting that Composite Activity to some extent measures users' influence and degree of social media usage. In terms of scientific activity indicators, the number of publications (P) and the number of citations (C) have a strong positive correlation, indicating that scholars with a higher number of publications also tend to have a higher citation count.

Variance Inflation Factor (VIF) calculations indicate that although FGR and CA have high VIF values, suggesting some overlap in measuring social media influence and potential multicollinearity, it is important to note that follower growth rate and activity level represent different aspects of social media accounts – status and behavior. Despite possibly sharing underlying logic, they describe different sides of scholars' social media accounts. Therefore, both variables are retained, with their VIF values within an acceptable range. Other indicators have low VIF values, indicating relative independence and no multicollinearity issues, as shown in Table 2. Overall, the average VIF value of the independent variables is 2.53, which is less than the reasonable range of 10, indicating that the regression model does not have multicollinearity issues and is suitable for regression analysis.

Table 2: Correlation Coefficient Matrix

|     | AD | TD | FGR | FR | CA | P | C | PC | AW | VIF |
|-----|-----|-----|-----|-----|-----|-----|-----|-----|-----|-----|
| AD  | 1 | -0.17079 | 0.03457 | 0.028453 | 0.073997 | 0.026723 | 0.000669 | -0.04026 | -0.05118 | 1.066425 |
| TD  | -0.17079 | 1 | 0.186552 | -0.04193 | 0.225964 | 0.008965 | 0.00315 | -0.02349 | -0.01586 | 1.105687 |
| FGR | 0.03457 | 0.186552 | 1 | -0.05172 | 0.89326 | 0.00785 | 0.007405 | 0.027709 | -0.02475 | 5.217847 |
| FR  | 0.028453 | -0.04193 | -0.05172 | 1 | -0.0479 | -0.02813 | -0.00849 | -0.026 | -0.03705 | 1.013585 |
| CA  | 0.073997 | 0.225964 | 0.89326 | -0.0479 | 1 | 0.059563 | 0.052463 | 0.049323 | -0.01751 | 5.413212 |
| P   | 0.026723 | 0.008965 | 0.00785 | -0.02813 | 0.059563 | 1 | 0.741161 | 0.24488 | 0.161994 | 2.603644 |
| C   | 0.000669 | 0.00315 | 0.007405 | -0.00849 | 0.052463 | 0.741161 | 1 | 0.602604 | 0.228269 | 3.870545 |
| PC  | -0.04026 | -0.02349 | 0.027709 | -0.026 | 0.049323 | 0.24488 | 0.602604 | 1 | 0.255006 | 1.902570 |
| AW  | -0.05118 | -0.01586 | -0.02475 | -0.03705 | -0.01751 | 0.161994 | 0.228269 | 0.255006 | 1 | 1.110207 |

**4.2 Regression Analysis and Model Construction**

In constructing the logistic regression model, the first step involves data preparation. This study considers the status of scientific social stratification mobility as the dependent variable, with social media indicators and scientific activity indicators as independent variables for logistic regression analysis. Prior to analysis, the random_state parameter is used to maintain consistency in the data split. The dataset is randomly divided to ensure each subset represents the characteristics of the original data, enhancing the accuracy of model evaluation. In the logistic regression analysis, data is split into a training set (70% of the original data) and a validation set (30% of the original data), with 321 observations in the training set. This split aids in evaluating model performance and generalizability. The dataset's distribution is random, ensuring accurate assessment. The sample distribution for the categories 'Increase' and 'Decrease' consists of 226 and 233 samples, respectively, suitable for logistic regression. The model successfully converges after 1000 iterations, improving overall classification performance.

Initially, a baseline logistic regression model is established using all available independent variables. In the logistic regression analysis, social media indicators and scientific activity indicators are analyzed, with variable selection based on the maximum likelihood estimation method to determine the inclusion of each candidate variable. Multiple logistic regression models are constructed based on the results of feature selection. These models' performance is assessed and compared using cross-validation techniques, focusing on metrics such as accuracy, AUC, and log-likelihood. Coefficients and exponentiated coefficients in each model are interpreted to understand the impact of independent variables on the dependent variable.

This chapter aims to study the stratification and mobility of scientists in the era of social media, exploring the impact and extent of change in scientific social stratification due to social media. Research variables include social media account days (AD), post density, follower growth rate, following ratio, Composite Activity (social media user profile indicators), and the number of publications, citations, average citations per publication (scientific activity indicators), used as independent variables to analyze their impact on the dependent variable, i.e., the degree of change in scientific social stratification. The paper comprehensively analyzes the significance and direction of multiple independent variables' impact on scientific social stratification, exploring the determinants of scientific social stratification. According to the regression results, the overall model interpretation is as follows: Overall, as shown in Table 3, the follower growth rate (FGR) shows a significant negative impact, while the following ratio (FR), Composite Activity (CA), number of publications (P), and journal authorship weight (AW) exhibit positive impacts. The impact of post density (TD), number of citations (C), and average citations per publication (PC) is statistically less significant, with the influence of PC being almost negligible.

Table 3: Logistic Regression Analysis of Scientific Social Stratification

|  | Variables | Coef. | Std.Err. | z | P>|z| | Exp(B) | Wald | VIF |
|---|---|---|---|---|---|---|---|---|
| Social Media Indicators | AD | -0.000101 | 0.000148 | -0.682705 | 0.494793 | 0.999899 | 0.465714 | 1.066425 |
|  | TD | 0.032025 | 0.022440 | 1.427148 | 0.153537 | 1.032544 | 2.036726 | 1.105687 |
|  | FGR | -13.661805 | 2.640962 | -5.173041 | 0.000000 | 0.000001 | 26.760354 | 5.217847 |
|  | FR | 0.266868 | 0.088335 | 3.021089 | 0.002519 | 1.305868 | 9.126986 | 1.013585 |
|  | CA | 0.019779 | 0.006246 | 3.166435 | 0.001543 | 1.019976 | 13.012936 | 5.413212 |
| Scientific Activity Indicators | P | 0.105302 | 0.029191 | 3.607337 | 0.000309 | 1.111046 | 13.012936 | 2.603644 |
|  | C | 0.003807 | 0.001905 | 1.998160 | 0.045699 | 1.003814 | 3.993703 | 3.870545 |
|  | PC | 0.004191 | 0.015072 | 0.278089 | 0.780944 | 1.004200 | 0.077320 | 1.902570 |
|  | AW | 0.365668 | 0.094083 | 3.886651 | 0.000102 | 1.441476 | 15.106074 | 1.110207 |

Commonly, it's believed that joining social media platforms earlier is advantageous for users. Through logistic regression analysis, the impact of the length of time spent on social media platforms (Account Days) on the status of scientific social stratification mobility (TAG1_binary) is examined. The analysis shows that the coefficient of Account Days is -0.000101 with a standard error of 0.000148. The z-value is -0.682705, and the P-value is 0.494793, neither reaching the standard for statistical significance. This suggests that while the duration of social media use might be a factor influencing scientists' social stratification, the time of joining social media platforms doesn't directly indicate its impact on scientific social stratification. In this study, the earlier a scientist joins a social media platform, it does not necessarily mean they will significantly impact the mobility of scientific social stratification. Further analysis is required.

The logistic regression analysis of the impact of post density (Tweet Density), an important indicator on social media platforms, on scientific social stratification, shows a coefficient of 0.032025. Although there is a positive relationship between post density and scientific social stratification, it is not statistically significant (P-value > 0.05). The logistic regression P-value for the Followers Growth Rate (FGR) indicates statistical significance in its impact on scientific social stratification. The coefficient of FGR is -13.661805, showing a significant negative impact on scientific social stratification. An Exp(B) value close to 0 indicates that an increase in the follower growth rate significantly reduces the likelihood of scientific social stratification. This, combined with the physical meaning of FGR, suggests that when scholars invest significant effort in social media, it may negatively affect their level of scientific social stratification.

Hypothesis 1 posits that a proactive attitude in social media use may play a significant role. Scientists who are more willing to connect with other scholars are expected to benefit more from social networks than those with inactive accounts. Therefore, it's hypothesized that an active behavioral character significantly and meaningfully impacts one's position in scientific social stratification. From the regression analysis results, we find that the logistic regression outcome for the Following Ratio (FR) is significant (p-value = 0.002519) and positive. Considering that the Following Ratio can reflect a user's strategy on social media, a higher FR might improve one's status in scientific social stratification, suggesting that actively establishing connections could bring unexpected benefits. The Exp(B) value of 1.305868 implies that with an increase in FR, the likelihood of advancing in scientific social stratification significantly increases. Therefore, the null hypothesis H0 is rejected, supporting the alternative hypothesis that active social media behavior significantly impacts a scientist's position in scientific social stratification. This finding underscores the importance of active interaction and networking on social media for enhancing a scientist's status in the scientific community.

Hypothesis 2 explores the correlation between scientists' activity level on social media platforms and their status in scientific social stratification. Logistic regression analysis shows that the coefficient for Composite Activity is 0.019779, with a standard error of 0.006246, z-value of 3.166435, and P>|z| value of 0.001543, well below the significance threshold of 0.05. This indicates a significant positive correlation between Composite Activity and scientific social stratification. An Exp(B) value of 1.019976 suggests that with each unit increase in Composite Activity, the odds of moving up in scientific social stratification increase by approximately 1.02 times. These results support the alternative hypothesis H3, indicating a significant correlation between a scientist's activity level on social media platforms and their status in scientific social stratification. Specifically, the more active scientists are on social media, the higher their position tends to be in scientific social stratification. However, considering the significant negative correlation presented by FGR, it can be concluded that while a moderate level of social media activity is necessary, excessive investment in social media platforms can be counterproductive.

Additionally, this study also examines the impact of scientific activity indicators, including the number of published papers (Publications, P), the number of citations a scientist receives (Citations, C), the average number of citations per paper (Per Cited, PC), and the weight of journal authorship (Amount Weight, AW) on scientific social stratification.

Logistic regression analysis reveals that the number of published papers has a significant positive effect on scientific social stratification, with a coefficient of 0.105302, an Exp(B) value of 1.111046, and a p-value of 0.000309. An Exp(B) value of 1.111046 implies that an increase in the number of published papers by a scientist correspondingly increases their chances of advancing in scientific social stratification. The number of citations, C, has a marginally significant impact on scientific social stratification, with a coefficient of 0.003807, an Exp(B) value of 1.003814, and a p-value of 0.045699. However, in this model, the average number of citations per paper, PC, does not show a statistically significant impact on scientific social stratification. Although the average number of citations is an important metric for evaluating academic achievements, it may not directly influence a scientist's position in scientific social stratification. Therefore, the average number of citations per paper might have a weaker correlation with scientific social stratification. Finally, the coefficient for journal authorship weight (AmountWeight) is 0.365668, indicating that for every additional unit increase in the frequency of journal appearances, the log odds of being in a higher tier of scientific social stratification significantly increase. The p-value of 0.000102 signifies that the frequency of journal appearances has a significant and positive effect on scientific social stratification. It's important to consider that this finding might be influenced by the limitations of data sources.

### 4.3 Logistic Regression Model Validation and Evaluation

In the process of constructing logistic regression models, selecting appropriate model variables is crucial. Variable selection is not only a technical issue but also involves a balance between model interpretability and practicality. A good model should be accurate in prediction and possess explanatory power, and variable selection must also consider the specific research context. This paper aims to explore how to improve Logistic regression models and to explain the concepts of full and sub-models, transitioning from a full model to a sub-model in a recursive manner to identify the optimal model. The full model includes all the independent variables involved in the experimental design, while sub-models encompass all possible models containing all available independent variables. For constructing sub-models, a systematic exploration of every possible combination of independent variables is conducted using a random combination method. This method involves randomly combining variables and assessing their impact on model performance. In this process, function values are calculated to evaluate the contribution of each combination to the model's performance. Typically, lower function values indicate better data fitting by the model. This approach allows for a comprehensive examination of various variable combinations, enabling the identification of the most effective model configuration. By considering a wide range of potential variable combinations, the final model chosen is more likely to be both predictive and relevant to the research question at hand. This method ensures that the final logistic regression model is not only statistically robust but also contextually relevant and interpretable.

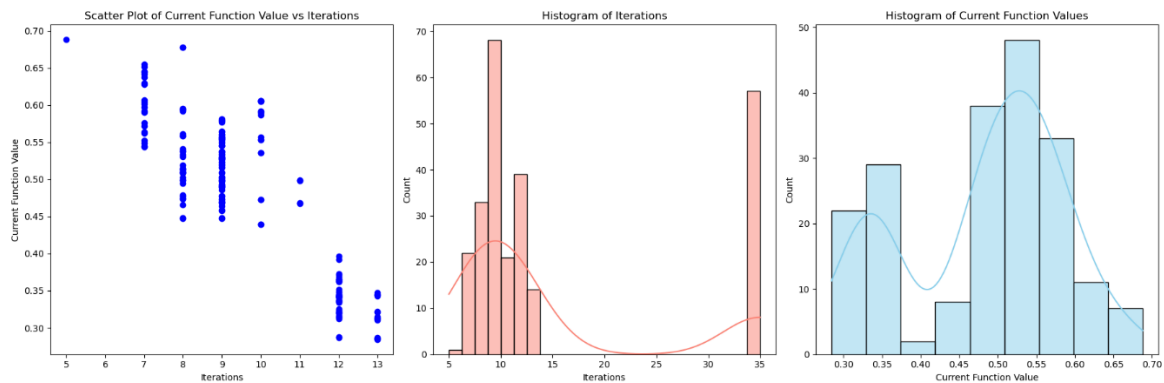

Figure 4: Logistic regression model validation and evaluation

As shown in figure 5, in terms of the stability of the model with respect to the data, the scatter plot of scatter plot of convergence behavior in Logistic Regression Models reveals a concentration of current function values within a specific range of iteration counts. This indicates that most models converge within a relatively small number of iterations, while only a few require a higher number of iterations to reach convergence. The minimal fluctuation in current function values suggests that the model fitting is relatively stable with respect to the data. The histogram of distribution of Iteration counts for convergence in Logistic Regression Models displays the distribution of iteration counts required for model convergence. The majority of model instances converge after less than 10 iterations. However, there are a few models that do not converge even after 35 iterations, suggesting potential issues in fitting these data points or inherent characteristics of the data making it challenging for the model to converge. In the histogram of distribution of function values in Logistic Regression Model displays the distribution of current function values demonstrates a tendency towards central concentration, but with some longer tails, indicating that some model instances have lower fitting quality than others. Lower function values indicate better model fits, while higher values may suggest that some models have not fitted the data effectively. This variation in function values implies a need for careful consideration of model suitability and potential refinement to ensure more consistent and accurate fitting across different data instances.

On the basis of the random combination method for model selection, coupled with the dual verification of stepwise regression using the backward elimination method, the model is simplified from the full variable model. This process begins by calculating the outcomes of models with all possible combinations of independent variables. Then, in the randomly combined models, variables are eliminated one by one. In the outcomes of the full variable combination models, variables that meet the significance level of the test probability P are selected and retained, while the least effective variables are eliminated to derive the required equation. The model is thus simplified from the full variable basis. In the process of model selection, factors such as predictive accuracy and model complexity are considered. The optimal combination of variables is ultimately chosen based on their physical meaning, thereby constructing the final optimized model. Stepwise regression, as a traditional method of variable selection, offers a way to balance model complexity and predictive accuracy in practice. Finally, by comparing the performance of different models, the most suitable model for specific data and predictive objectives is determined. When using stepwise regression methods, caution must be exercised to avoid potential statistical pitfalls, considering the interrelations between variables and potential interaction effects. This method might sometimes overlook variables that are not statistically significant but theoretically important. The process outlined above aids in constructing an efficient and reliable multivariate logistic regression model, providing a solid foundation for data-driven decision-making. This approach ensures that the final logistic regression model is not only efficient in terms of variable selection but also robust and reliable in its predictions, thereby enhancing the overall quality of research findings and their applicability in real-world scenarios.

Table 4: Performance of Different Models

|  | Variables | Model1 | Model2 | Model4 | Model7 | Model10 | Modwl14 |
|---|---|---|---|---|---|---|---|
| Social Media Indicators | AD | -0.000101 | | | | | |
|  | TD | 0.032025 | 0.034417 | | | | |
|  | FGR | -13.661805*** | -13.597224*** | -13.235143*** | -10.481163*** | -9.9687460*** | -10.874992*** |
|  | FR | 0.266868** | 0.269170** | 0.270392** | 0.241053** | 0.206699** | |
|  | CA | 0.019779** | 0.019505** | 0.019458** | 0.015098** | 0.014670* | 0.014174* |
| Scientific Activity Indicators | P | 0.105302*** | 0.103881*** | 0.101555*** | 0.136243*** | 0.152015*** | 0.147756*** |
|  | C | 0.003807* | 0.003850* | 0.003955** | | | |
|  | PC | 0.004191 | 0.003367 | | 0.033523** | | |
|  | AW | 0.365668*** | 0.365376*** | 0.364572*** | | | |
| Model Performance | Accuracy | 0.804 | 0.789, | 0.789 | 0.804 | 0.819 | 0.831 |
|  | Precision | 0.895 | 0.891 | 0.891 | 0.959 | 0.943 | 0.962 |
|  | Recall | 0.708 | 0.680 | 0.681 | 0.653 | 0.694 | 0.6944 |
|  | F1-Score | 0.791 | 0.680 | 0.772 | 0.777 | 0.8 | 0.806 |
|  | Log-Likelihood | -91.334 | -91.568 | -92.242 | -107.720 | -119.591 | -127.156 |

| Variables | Model1 | Model2 | Model4 | Model7 | Model10 | Modwl14 |
|---|---|---|---|---|---|---|
| LL-Null | -222.237 | -222.237 | -222.237 | -222.237 | -222.237 | -222.237 |
| Pseudo R-squared | 0.589 | 0.588 | 0.585 | 0.515 | 0.462 | 0.428 |
| LLR p-value | 3.1985672439685 67e-51 | 6.786723103 141654e-52 | 3.0019033751 889203e-53 | 1.723577955 160804e-47 | 2.7340302278 146836e-43 | 5.632566622 899111e-41 |
| AIC | 202.668 | 201.136 | 198.484 | 227.441 | 249.181 | 262.312 |
| BIC | 240.383 | 235.079 | 224.884 | 250.069 | 268.038 | 277.398 |

  The overall effectiveness of the model in predicting the direction of change in scientific social stratification (i.e., increase or decrease) is relatively good. Particularly in predicting category 0 (Decrease), the model achieves a recall rate of 86%, indicating its proficiency in identifying instances of decreasing scientific social stratification. However, for category 1 (Increase), despite a high precision of 83%, the recall rate is lower at 62%. The confusion matrix shows that for label 0, there are 55 correct predictions and 11 false positives; for label 1, there are 49 correct predictions and 23 false positives. The model is slightly more adept at predicting decreases in scientific social stratification (label 0) than increases (label 1). To further understand the impact of different features - social media indicators and scientific activity indicators - on the stratification and mobility within the scientific community, multiple model cross-validation is conducted. Different regression models are constructed based on these features, analyzing the relationship between these variables and the stratification and mobility in the scientific community. The most important feature variables are selected through stepwise regression to enhance the model's interpretability. Models 2, 4, 7, 10, and 14 are chosen for inclusion. The optimal model, Model 4, is identified with an AIC (Akaike Information Criterion) value of 198.484. This approach to model selection and validation, incorporating both precision and recall metrics, ensures a balanced evaluation of the model's predictive capabilities. By focusing on both the accuracy of predictions (precision) and the model's ability to capture relevant cases (recall), a more comprehensive understanding of the model's performance is achieved. The chosen optimal model, with its lower AIC value, indicates a good balance between model complexity and fit, making it a robust tool for analyzing and predicting changes in the stratification and mobility within the scientific community.

  In the stepwise regression process from the full model to the sub-model, balancing factors such as predictive accuracy and model complexity is crucial. Among these factors, the Akaike Information Criterion (AIC) value is a key consideration, as the full model, with its more comprehensive perspective due to the inclusion of more variables, typically benefits parameter estimation accuracy. In the stepwise regression analysis, the overall model is first considered, and its AIC is calculated. Then, the process involves removing one variable from the complete model and recalculating the AIC for these new models, with the aim of selecting the model with the smallest AIC as the current best model. When selecting a model, significant influence of independent variables on the dependent variable is closely monitored, indicated by their corresponding Sig values being below 0.05. Moreover, the best model is identified by comparing different models' accuracy, AUC, and log-likelihood. Particularly in handling imbalanced class data, measures including precision, recall, and F1 score are also considered to thoroughly evaluate model performance. However, it's important to note that in regression analysis where the dependent variable is binary, it can be challenging for independent variables to produce significant differences. Changes in independent variables may not significantly affect the dependent variable. Thus, even if the overall regression model is significant, and independent variables show non-significant P-values, the model can still hold considerable practical value, especially if it presents a high R-squared value and a high rate of correct classification. This applies particularly in the context of overall models 1 and 2, indicating the models' relevance despite the apparent non-significance of individual independent variables.

This study employs cross-validation, ROC curves, and other methods to evaluate the performance of the model. Such methods not only improve the model's predictive accuracy but also enhance its interpretability, offering new perspectives for handling more complex datasets. We conducted an in-depth comparison of logistic regression models with other common prediction models, such as decision trees, random forests, and support vector machines (SVM). As illustrated in Figure 5, a visual representation of the performance comparison of different models on the test set is presented. By applying these models to the same dataset and assessing their performance using cross-validation methods, we gain a comprehensive understanding of each model's strengths and limitations. The chart includes the AUC scores of the following models: Logistic Regression, Decision Tree, Random Forest, and SVM, with each model's performance indicator represented by a bar graph in different colors.

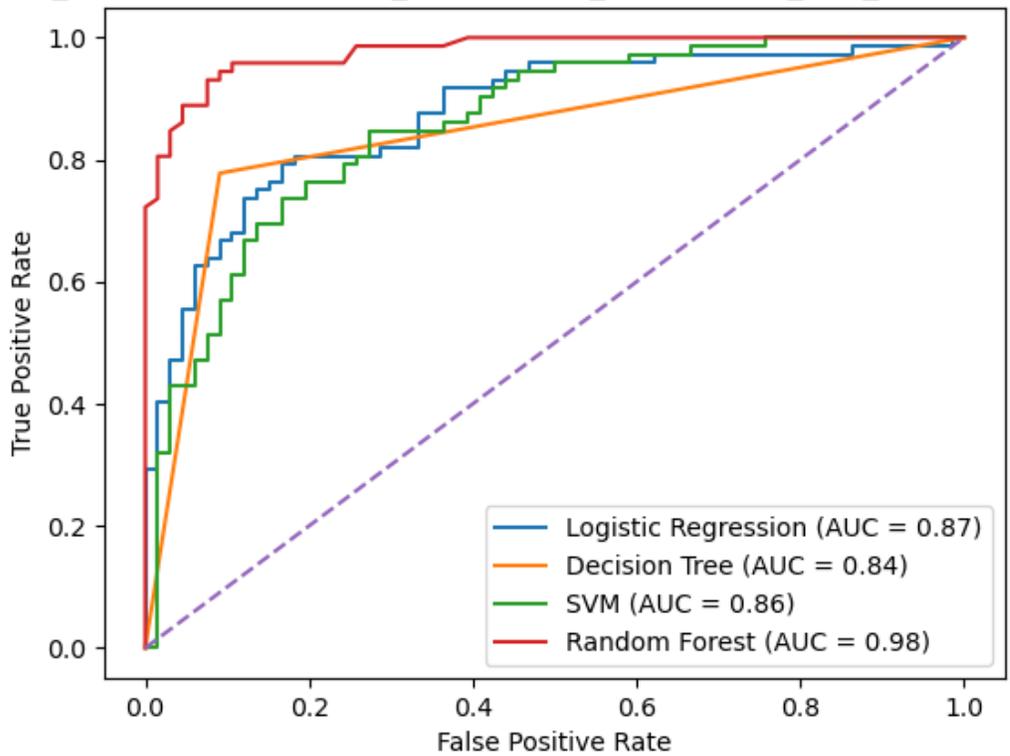

Figure 5: Logistic regression model validation and evaluation

To conduct robustness tests, we first utilize Receiver Operating Characteristic (ROC) curves to visually represent the performance of each model. ROC curves are tools for evaluating the performance of classification models, depicting the relationship between the True Positive Rate (TPR) and the False Positive Rate (FPR), demonstrating the model's ability to distinguish between classes. We further calculated the Area Under the Curve (AUC) score for each model as a quantifiable performance indicator. Overall, the Random Forest model demonstrated the best performance among these four, with a higher AUC score indicating stronger classification capabilities. Additionally, our study includes robustness tests to ensure the reliability and effectiveness of the models. The Logistic Regression model, in particular scenarios, exhibited superior interpretability and flexibility. Although its performance might not match that of the Random Forest model, Logistic Regression still holds unique advantages in handling linear relationships and providing interpretive insights into the model. In summary, this comprehensive comparative approach not only enhances the predictive accuracy of the model but also its interpretability, enabling us to develop a powerful and effective model tailored to the specific requirements and characteristics of the dataset at hand. By identifying key variable combinations that affect the model's predictive capabilities, we offer new perspectives and methods for handling more complex datasets.

**4.4 Analysis of Influencing Factors in Scientific Social Stratification and Mobility**

This paper utilizes SHAP (SHapley Additive exPlanations) to visualize and analyze the impact of different features within the optimized logistic regression model, evaluating the contribution of variables to the model and the importance of various factors affecting scientific social stratification and mobility. The importance ranking of independent variables is shown in Figure 6. This study explores the impact of social media on scientific social stratification and mobility, dividing the independent feature variables into two main categories: social media indicators and scientific activity indicators. The bar graph in the figure demonstrates the importance of each feature, with social media indicators including AD, TD, FGR, FR, Composite Activity (CA), and scientific activity indicators comprising Publications, Cite, PerCited, AmountWeight. Arranged vertically by importance, the bar graph quantitatively analyzes the feature importance in the logistic regression model using the mean absolute SHAP values. The horizontal axis lists the mean absolute SHAP values as a measure of each feature's average impact on model predictions. Notably, "FGR" exhibits the most pronounced average effect, with a mean absolute SHAP value of +0.26, highlighting its significant role in the model's decision-making process. "Cite" also has a notable impact, with a mean absolute SHAP value of +0.2. "Publications" and "AmountWeight" have a moderate influence, with mean absolute SHAP values of +0.05 and +0.04, respectively, indicating consistent but smaller effects. Features like "AD," "Composite Activity (CA)," "PerCited," "TD," and "FR" show more moderate impacts, with mean absolute SHAP values ranging around +0.01 to +0.02. The uniform red color of the bars indicates the unidirectional influence of feature values on the prediction outcome, with larger values generally contributing to an increase in the model's output. This visualization assists in understanding the relative importance of features within the predictive framework, which is crucial for affirming the model's interpretability in an academic context.

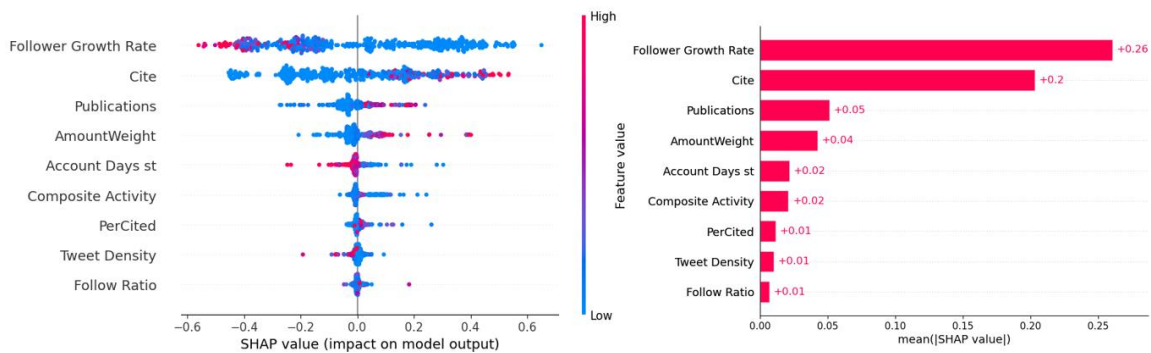

Figure 6: Logistic regression model validation and evaluation

Each point on the SHAP summary plot represents the SHAP value of a feature in an individual prediction. The colors signify feature values (blue for high, pink for low). Features are vertically stacked with the most influential ones at the top. The horizontal position of the points indicates their impact on the model output, with negative values suggesting a decrease and positive values indicating an increase in the prediction outcome. The SHAP summary plot reveals the contribution of each feature to the predictive model output. High feature values are displayed in blue, while low values are shown in pink, with the color intensity representing the magnitude of the feature's value. The x-axis displays the SHAP values, quantifying each feature's impact on the model's predictions. SHAP values close to zero suggest minimal impact on model results, while values far from zero indicate a greater influence. The most impactful features are "Followers Growth Rate" and "Citations," which exhibit a mix of positive and negative effects on the model output. "Publications" and "Amount Weight" have narrower distributions, suggesting more consistent impacts on predictions. This chart is crucial for interpreting the model, as it provides a visual representation of feature importance and their individual contributions to the prediction outcome.

In our study, we not only assessed the overall contributions of variables in the predictive models but also delved into the dynamic trends of these contributions. As illustrated in Figure 7, these LOWESS (Locally Weighted Scatterplot Smoothing) charts meticulously reveal the relationship between feature values and their corresponding SHAP values.

On one hand, key metrics measuring the impact of features on model predictions were explored, examining the extent of influence of independent variables on the model's predictive power, thereby enhancing the transparency and credibility of the predictive model. Through SHAP analysis, the magnitude and direction of each feature's impact were assessed, as well as the variability and consistency of this impact across different values. This detailed analysis aids in refining model features, improving prediction accuracy, and ensuring the robustness of model deployment, which is crucial for hypothesis testing, model validation, and elucidating the mechanisms underpinning complex logistic regression models.

On the other hand, we conducted a thorough comparative analysis using the LOWESS method to contrast the SHAP value trends of each variable in both the complete and optimized models across two logistic regression frameworks – the overall model and the optimized model. This comparison elucidated the differential impacts of model optimization on the predictive dynamics of shared variables, deepening our understanding of how improvements in the model influence the relationship between feature values and their impacts on model predictions. This comparative analysis allowed us to comprehend the dynamic behavior of each variable within different model frameworks and how these behaviors impact the accuracy and reliability of model predictions. It also played a key role in understanding the complex dynamics within the model and enhancing the interpretability of the discussed logistic regression models.

As shown in Figure 7, the composite chart presents a series of LOWESS trend analyses, elucidating the relationships between selected feature values and their SHAP values in two logistic regression models – the overall model and the optimized model. The LOWESS trend lines overlay the scatter of SHAP values, revealing the potential impact patterns of each feature. The "overall model" is indicated by solid lines, while the "optimized model" is represented by dashed lines. For instance, the "Followers Growth Rate" subplot shows a primary negative correlation in both models, with the optimized model exhibiting a more pronounced decline in SHAP values as growth rates increase. Conversely, "Composite Activity" and "Publications" display positive correlations, with SHAP values stabilizing as feature values increase, suggesting a diminishing incremental impact at higher levels of these features.

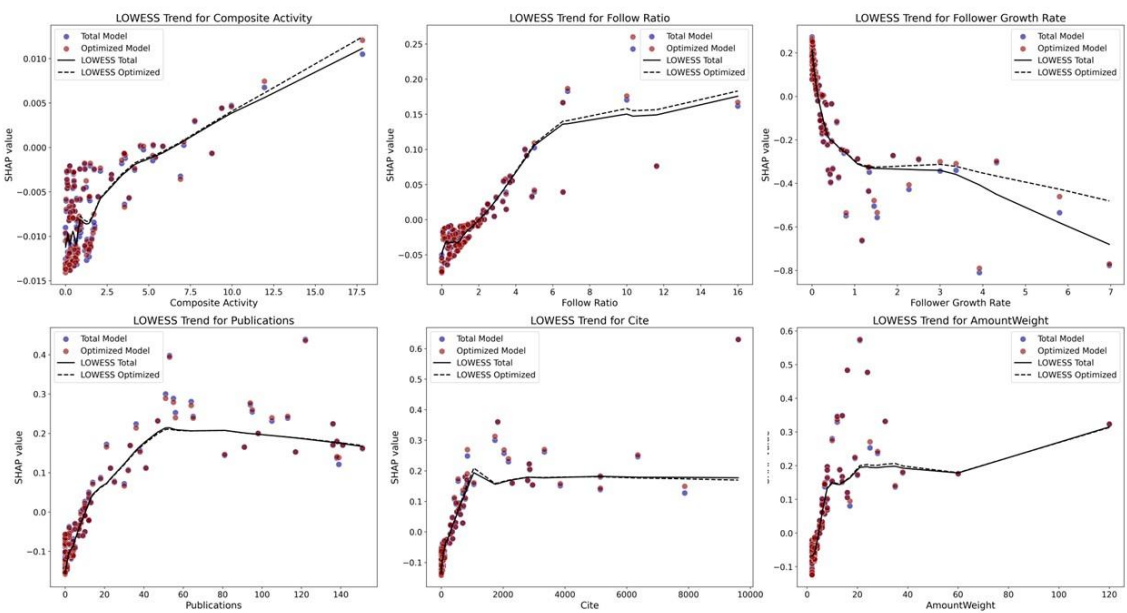

Figure 7: Comparative LOWESS Analysis of Feature Contributions in Logistic Regression Models

In our research, we delved deeply into the role of the "Composite Activity" (CA) variable in the predictive model, paying special attention to the trend of its SHAP values. The analysis showed that as the "CA" variable increased, its SHAP values exhibited an almost linear upward trajectory, a trend that was consistently observed in both types of models. This near-linear relationship suggests a direct and proportional link between "CA" and the model's predictive outcomes. Furthermore, observing the trajectory of the SHAP values for the "CA" variable, we found a strong positive correlation with the predictive outcomes. This finding was particularly significant in the process of model optimization, indicating that as "CA" increases, the magnitude of the model output matches its increase. This near-linear relationship not only reveals the importance of "CA" in the predictive model but also provides clear guidance for understanding and adjusting the role of the "CA" variable within the model.

The analysis of the "Follow Ratio" (FR) variable within the predictive model. Observing the changes in the SHAP values for the "FR" variable, we detected a significant trend: initially, an increase in "FR" exhibited a positive slope, which then transitioned to a plateau phase. This finding suggests that within the predictive model, there is an optimal range for "FR" values, where values within this range positively contribute to the model's output, but beyond a certain level, the effect tends to saturate. Moreover, this trend may also reflect the limited impact of social dynamics in the model's predictions. Specifically, this plateau effect could be the result of diminishing marginal utility of "FR" within the range of our observed data, indicating that beyond a certain point, additional increases in "FR" yield diminishing returns for the model's output. This discovery is crucial for understanding the role of social dynamic variables like "FR" in influencing model predictions and provides key insights for optimizing the application of such variables in the model.

The analysis showed a significant negative correlation between "Followers Growth Rate" and predictive results, clearly evidenced by the negative LOWESS trend in its SHAP values. Specifically, a higher "Followers Growth Rate" was accompanied by a decrease in model output, particularly pronounced in the optimized model. This finding overturns intuitive expectations that a higher growth rate of followers might correlate with a higher probability of the predicted event. This inverse relationship reveals a complex phenomenon where rapid growth in followers might inversely correlate with the desired outcomes in the model's application context. This counterintuitive result necessitates further research to explore its causes and underlying mechanisms. In summary, our study offers a new perspective on the influence of "Followers Growth Rate" in different predictive models, especially in understanding how it impacts model outputs in non-traditional ways. These findings not only challenge traditional understandings but also provide new directions for future research, particularly in exploring the relationship between social media dynamics and predictive models.

Analyzing the Impact of the ' Publications ' Variable in the Predictive Model. Observing the changes in its SHAP values, we identified a logarithmic trend: initially, the SHAP values for "Publications" showed an upward trajectory, but then reached a plateau stage. This trend indicates that while academic output (i.e., the number of publications) generally has a positive impact on predictive outcomes, this impact does not increase indefinitely, and there is a clear limit to its marginal effects. Additionally, this trend reinforces the concept of the nonlinear impact of academic output on model predictions. Specifically, before reaching a certain quantity, the number of publications has a significant positive impact on the model's predictions; however, beyond this threshold, its influence on predictive outcomes tends to stabilize. This finding is crucial for understanding how academic output influences model predictions at different levels and provides important theoretical support for considering the role of academic output in future optimizations of predictive models.

Analyzing the Impact of the 'Cite' Variable in the Predictive Model. The analysis revealed a complex non-linear relationship between this variable and the predictive output. Initially, an increase in "Cite" was accompanied by a rise in its SHAP values, indicating a positive trend. However, this relationship started to change after reaching a turning point at higher citation levels, where the rate of increase in SHAP values slowed and eventually stabilized, slightly decreasing at even higher values. This trend likely unveils a significant saturation effect, suggesting that beyond a certain threshold, additional citations start to diminish in their impact on model predictions. This finding echoes the law of diminishing marginal returns in economics, suggesting that in the predictive model, the contribution of "Cite" to the output no longer increases proportionally after reaching a certain point. This non-monotonicity indicates that the model's sensitivity to additional citations decreases for higher citation values, reflecting different treatment of various citation levels within the predictive model, which is significant for understanding and optimizing the variable's role in complex predictive models.

Analyzing the 'AmountWeight' Variable's Impact on Predictive Model Output. Our findings indicate a significant positive correlation between the SHAP values of this variable and its numerical value, a trend corroborated by the ascending trajectory in the LOWESS analysis. Specifically, the positive correlation between "AmountWeight" and the outcome variable suggests that as "AmountWeight" increases, so does its contribution to the model's predictive outcome.

# 5 CONCLUSION

This study resides at the intersection of Scientometrics and Sociology of Science, launching from scientometric theory towards a convergence with sociological research. Although there's a wealth of research on scientific social stratification to date, the majority focuses on the pre-social media era. The advent of social media has undeniably impacted human society, making the exploration of its influence on scientific social stratification in the digital age a valuable addition to the fields of Sociology of Science and Scientometrics. This exploration also represents an experimental foray of Scientometrics into the realm of Sociology. This paper investigates the impact of scientists' use of social media post-its emergence on the stratification and mobility within the scientific community in the digital age, uncovering the latent connections between academic status and social media activity. Building upon the theoretical framework of the Scientific Activity Theory model proposed in the author's prior research, this empirical study primarily seeks to validate whether the emergence of social media has facilitated mobility within scientific social stratification. Based on a sound experimental design and employing interpretable logistic attribution algorithms, we arrived at a significant conclusion: social media has indeed fostered stratification and mobility within the scientific community, to a certain extent. This finding not only statistically validates the hypothesis that the use of social media positively influences scientific social stratification but also elucidates, through interpretable SHAP values, how social media activity variables impact the state of stratification and mobility in the scientific community, addressing the 'how' aspect. The empirical results from the interpretable logistic attribution algorithm reveal that this promotional mechanism is complex and non-linear. In other words, social media activity within a certain range positively influences scientists' status in scientific social stratification. However, beyond a certain threshold, this influence turns negative, a conclusion that also aligns, to some extent, with our everyday experiences.

This paper innovatively utilizes an interpretable logistic attribution algorithm, a method that straddles both statistical methodology and machine learning algorithms, particularly effective for binary classification problems. In statistics, logistic regression is used for hypothesis testing and estimating the impact of variables. In machine learning, it is applied for prediction and model evaluation. Thus, the interpretable logistic regression algorithm emerges as a potent tool integrating machine learning with statistical analysis. Logistic regression also serves as a classifier within machine learning, and this research synergistically merges these two applications. Its unique position in both statistics and machine learning stems from its dual capability for statistical inference, such as hypothesis testing and estimating the influence of independent variables on the dependent variable, and as a predictive classification algorithm. This blend of inferential and predictive capacities is relatively rare among other algorithms like deep learning, random forests, or SVMs. Although these algorithms may surpass logistic regression in predictive performance, they are not typically suited for traditional statistical inference. Logistic regression stands out as one of the few algorithms that can address both the statistical significance of hypothesis testing and predictive modeling in machine learning, evaluating the impact of independent variables on the dependent variable while also functioning as a classifier. While other algorithms can also fulfill the needs of statistical inference and prediction to some extent, they usually focus more prominently or perform better in one aspect.

In the era of social media, the phenomena of stratification and mobility among scientific researchers not only reveal the dynamic structural changes within the scientific field but also reflect the distribution of power and resources within the scientific community system. Research on the scientific evaluation mechanism indicates that this mechanism not only directly impacts the personal career development of scientists but also constitutes a crucial factor in the healthy development of the scientific community system. Moreover, the stratification and mobility of researchers play a positive role in promoting the diversification of scientific knowledge and the formation of innovative thinking, contributing to the healthy development of the entire scientific community system.

The emergence of social media has opened new channels for academic influence, allowing scholars to engage in academic exchange, public science popularization, and influence dissemination via social media platforms, transcending the limitations of traditional academic publications. The dissemination and reception of academic achievements on social media are influenced by factors different from those of traditional academic evaluation systems, such as content accessibility, topic popularity, and social network structure. Interactive behaviors like likes, shares, and comments have become new metrics for measuring the social impact of academic work. The differences between these new metrics and traditional academic achievement indicators have led to changes in the stratification of the scientific community. The rise of social media and the reshaping of academic status, the impact of social media on the hierarchical structure of academia, specifically the influence of social media on academic recognition, and the exploration of the network extension of academic influence, demonstrate that the immediacy and extensive coverage of social media enable some scholars to achieve influence on these platforms that may surpass their traditional academic status.

- In principle, scholars with high h-index typically enjoy higher academic status. It is inferred that the academic influence of these scholars may extend into the network, manifesting as higher rankings online. Therefore, we can speculate that in the network, scholars with high academic influence in the real world also generally possess higher network rankings. However, this relationship may not be absolute and could be influenced by factors such as the scholar's proficiency in using social media, the presentation of content, target audience, etc. The research background and academic achievements of high h-index scholars are recognized and reinforced in social media and online spaces. Their expertise and reputation might garner them more attention on social media, further enhancing their academic status.

- Social media challenges the traditional academic hierarchy. Scholars active on social media, especially those with lower status in the traditional academic world, may gain broader public attention and influence. Additionally, social media provides a new platform for young researchers and scholars in marginal disciplines to showcase their research and build influence, helping to break down traditional academic barriers and promote more diverse and equitable academic exchanges. Overall, social media has redefined the standards for measuring academic status and influence to some extent, profoundly impacting the stratification of the scientific community. This change brings both opportunities and challenges to academia, particularly in evaluating and recognizing the value of academic work. Therefore, it is crucial for all members of the academic community to understand and adapt to this change.

Research Limitations: Firstly, in terms of experimental design, using the percentage rank of a scientist's H-index and the percentage rank of followers on academic social media as alternative indicators to represent the levels of scientific social stratification before and after the advent of social media has certain limitations. However, based on a deep interpretation of the theory of scientific social stratification and an in-depth analysis of its mechanisms, using these as surrogate measures for stratification based on scientific recognition is reasonably justifiable. Scientific recognition acts as a catalyst in the functioning of the contemporary scientific community system. Additionally, the selection of the field of scientometrics, although representative, is somewhat narrow. Future research could consider a broader range of scientific fields or comparisons between different disciplines. In terms of indicator selection, while the chosen indicators are representative, some may be overlooked. As a data-driven study, the research design is initially limited to a certain extent by data structure. Therefore, future studies could consider revisiting the research topic at a theoretical level, constructing a system of indicators for scientific social stratification, enhancing the research design, and enriching the data to solidify the research. Future Research Directions: Considering the impact and limitations of the evaluation indicator system, it is necessary to consider more dimensions. Future research might need to include a wider range of indicators to gain a deeper understanding of these relationships. It should build an indicator system for scientific social stratification from the perspective of scientific activity theory and account for the heterogeneity within academia and differences between disciplines. Moving from a single indicator to multiple indicators can more comprehensively depict the level of a scholar's scientific social stratification.

**Author Statement**

This study represents an independent research endeavor, built upon extensive accumulated expertise and insights within the fields of Scientometrics and Sociology of Science. It exhibits a high degree of originality and innovation in research design, data sources, and analytical methods.

**Acknowledgements:**

I am deeply grateful to Professor Wang Xianwen, for his invaluable guidance. It was he who led me into the field of Altimetric, opening up a whole new world for me. His mentorship goes beyond academic research; it has been a journey of companionship and guidance on my scholarly path, a treasure I will cherish for a lifetime. Additionally, I am profoundly honored and thankful to Mr. Jonathan R. Cole for his recognition and support of my research topic selection. This acknowledgment has been incredibly gratifying and encouraging.

**REFERENCES**

[1] Yue, Y, Xianwen, W. (2021). Altmetrics Review over 2010-2020. Journal of Intelligence, (11),136-146. Doi:10.3969/j.issn.1002-1965.2021.11.020

[2] Cole, Jonathan R., and Stephen Cole. Social Stratification in Science. Chicago: University of Chicago Press, 1973.

[3] Zuckerman, H. (1970). Stratification in American science. Sociological Inquiry, 40(2), 235-257.

[4] Deville, P., Wang, D., Sinatra, R., Song, C., Blondel, V. D., & Barabási, A. L. (2014). Career on the move: Geography, stratification and scientific impact. Scientific reports, 4(1), 4770.

[5] Jones, B. F., Wuchty, S., & Uzzi, B. (2008). Multi-university research teams: Shifting impact, geography, and stratification in science. science, 322(5905), 1259-1262.

[6] Hargens, L. L., & Felmlee, D. H. (1984). Structural determinants of stratification in science. American sociological review, 685-697.

[7] Blank, G. (2013). Who creates content? Stratification and content creation on the Internet. Information, Communication & Society, 16(4), 590-612.

[8] Senanayake U, Piraveenan M, Zomaya A. The Pagerank-Index: Going beyond Citation Counts in Quantifying Scientific Impact of Researchers. Plos one. 2015;10(8):e0134794. DOI: 10.1371/journal.pone.0134794. PMID: 26288312; PMCID: PMC4545754.

[9] Haustein, S., Costas, R., & Larivière, V. (2015). Characterizing social media metrics of scholarly papers: The effect of document properties and collaboration patterns. PloS one, 10(3), e0120495.

[10] Pulido, C. M., Redondo-Sama, G., Sordé-Martí, T., & Flecha, R. (2018). Social impact in social media: A new method to evaluate the social impact of research. PloS one, 13(8), e0203117.

[11] Wilson, K. R., Wallin, J. S., & Reiser, C. (2003). Social stratification and the digital divide. Social Science Computer Review, 21(2), 133-143.

[12] Jeon, J. (2019). Rethinking scientific habitus: Toward a theory of embodiment, institutions, and stratification of science. Engaging Science, Technology, and Society, 5, 160-172.

[13] Bornmann, L. (2014). Do altmetrics point to the broader impact of research? An overview of benefits and disadvantages of altmetrics. Journal of informetrics, 8(4), 895-903.

[14] Kwiek, M. (2018). Changing European academics: A comparative study of social stratification, work patterns and research productivity. Routledge.

[15] Yuval-Davis, N. (2016). Beyond the recognition and re-distribution dichotomy: Intersectionality and stratification. In Framing intersectionality (pp. 155-169). Routledge.

[16] Hofstra, B., Kulkarni, V. V., Munoz-Najar Galvez, S., He, B., Jurafsky, D., & McFarland, D. A. (2020). The diversity–innovation paradox in science. Proceedings of the National Academy of Sciences, 117(17), 9284-9291.

[17] Davis, K., & Moore, W. E. (2019). Some principles of stratification. In Social Stratification, Class, Race, and Gender in Sociological Perspective, Second Edition (pp. 55-63). Routledge.

[18] Bendix, R. (1952). Social stratification and political power. American Political Science Review, 46(2), 357-375.

[19] Collins, R. (2019). The credential society: An historical sociology of education and stratification. Columbia University Press.

[20] Erdt, M., Nagarajan, A., Sin, S. C. J., & Theng, Y. L. (2016). Altmetrics: an analysis of the state-of-the-art in measuring research impact on social media. Scientometrics, 109, 1117-1166.

[21] Massey, D. S. (2004). Segregation and stratification: A biosocial perspective. Du Bois Review: Social Science Research on Race, 1(1), 7-25.

[22] Leite, F. (2016). The stratification of diversity: measuring the hierarchy of Brazilian political science. Brazilian Political Science Review, 10.

[23] Shirahase, S. (2010). Japan as a stratified society: With a focus on class identification. Social Science Japan Journal, 13(1), 31-52.

[24] Thye, S. R., & Harrell, A. (2017). The status value theory of power and mechanisms of micro stratification: Theory and new experimental evidence. Social Science Research, 63, 54-66.

[25] Mare, R. D. (2014). Multigenerational aspects of social stratification: Issues for further research. Research in Social Stratification and Mobility, 35, 121-128.

[26] Róbert, P. (2010). Stratification and social mobility. Handbook of European Societies: Social Transformations in the 21st Century, 499-536.

[27] Pearl, J. (2011). Principal stratification--a goal or a tool?. The international journal of biostatistics, 7(1), 1-13.

[28] McNamee, S. J., & Willis, C. L. (1994). Stratification in science: A comparison of publication patterns in four disciplines. Knowledge, 15(4), 396-416.

[29] Jalali, Z. S., Introne, J., & Soundarajan, S. (2023). Social stratification in networks: insights from co-authorship networks. Journal of the Royal Society Interface, 20(198), 20220555.

[30] Jeon, J. (2019). Rethinking scientific habitus: Toward a theory of embodiment, institutions, and stratification of science. Engaging Science, Technology, and Society, 5, 160-172.

[31] Gasparyan AY, Yessirkepov M, Duisenova A, et al. Researcher and Author Impact Metrics: Variety, Value, and Context. Journal of Korean Medical Science. 2018 Apr;33(18):e139. DOI: 10.3346/jkms.2018.33.e139. PMID: 29713258; PMCID: PMC5920127.

[32] Lincoln, A. E., Pincus, S., Koster, J. B., & Leboy, P. S. (2012). The Matilda Effect in science: Awards and prizes in the US, 1990s and 2000s. Social studies of science, 42(2), 307-320.

[33] McGinnis, R., & Long, J. S. (2014). 2 Entry into Academia: Effects of Stratification, Geography and Ecology. In The Academic Profession (pp. 342-365). Routledge.

[34] Nguyen, T. V., & Pham, L. T. (2011). Scientific output and its relationship to knowledge economy: an analysis of ASEAN countries. Scientometrics, 89(1), 107-117.
29

31[66] Collins, R. (2013). Some comparative principles of educational stratification. In Sociological Worlds (pp. 330-345). Routledge.

[67] Yeung, W. J. J. (2013). Higher education expansion and social stratification in China. Chinese Sociological Review, 45(4), 54-80.

[68] Davies, S., & Zarifa, D. (2012). The stratification of universities: Structural inequality in Canada and the United States. Research in Social Stratification and Mobility, 30(2), 143-158.

[69] Marginson, S. (2016). Global stratification in higher education. Higher education, stratification, and workforce development: Competitive advantage in Europe, the US, and Canada, 13-34.

[70] Chinn, S., Hasell, A., Roden, J., & Zichettella, B. (2024). Threatening experts: Correlates of viewing scientists as a social threat. Public Understanding of Science, 33(1), 88-104.

[71] Falkenberg, R., & Fochler, M. (2024). Innovation in technology instead of thinking? Assetization and its epistemic consequences in academia. Science, Technology, & Human Values, 49(1), 105-130.

[72] Francis, B., Henderson, M., Godec, S., Watson, E., Archer, L., & Moote, J. (2023). An exploration of the impact of science stratification in the English school curriculum: the relationship between 'Double'and 'Triple'Science pathways and pupils' further study of science. Research Papers in Education, 1-23.

[73] Faria, I., Montalvan, A., Canizares, S., Martins, P. N., Weber, G. M., Kazimi, M., & Eckhoff, D. (2024). The power of partnership: exploring collaboration dynamics in US transplant research. The American Journal of Surgery, 227, 24-33.

[74] Kristensen, P. M. (2015). Revisiting the "American social science"—Mapping the geography of international relations. International studies perspectives, 16(3), 246-269.

[75] Hoekman, J., & Frenken, K. (2013). Proximity and stratification in European scientific research collaboration networks: a policy perspective. In The geography of networks and R&D collaborations (pp. 263-277). Cham: Springer International Publishing.

[76] Shu, F., Sugimoto, C. R., & Larivière, V. (2021). The institutionalized stratification of the Chinese higher education system. Quantitative Science Studies, 2(1), 327-334.

[77] Yue, Y. (2023). Unraveling Altmetrics Evolution through Scientific Activity Theory. Information studies: Theory & Application, (11),73-80.doi:10.16353/j.cnki.1000-7490.2023.11.010.

[78] Klauschen, F., Dippel, J., Keyl, P., Jurmeister, P., Bockmayr, M., Mock, A., ... & Müller, K. R. (2024). Toward explainable artificial intelligence for precision pathology. Annual Review of Pathology: Mechanisms of Disease, 19.